\documentclass[runningheads]{llncs}
\usepackage[T1]{fontenc}
\usepackage{graphicx}
\begin{document}
\raggedbottom
\title{Translating Ethical Frameworks into User-Centred Anti-Social Behaviour Interventions.}

\titlerunning{Translating Ethical Frameworks into HCI Interventions}
%
\author{Rachel Hill\inst{1}\orcidID{0009-0002-4971-615X} 988315@swansea.ac.uk\and
Tom Owen\inst{2}\orcidID{0000-0002-5150-0246}  \and
Julian Hough\inst{3}\orcidID{0000--0002-4345-6759}}
\authorrunning{R. Hill et al}
%
\institute{Swansea University, Swansea SA1 8EN, UK}
\maketitle             
This manuscript is an author-generated preprint of a paper accepted for publication in 
HCII 2026 (Springer CCIS). The final version may differ.
\begin{abstract}
In 2025 one million Anti-Social Behaviour (ASB) cases were recorded in England \& Wales, impacting community cohesion. Statutory guidance presents punitive interventions that lack technological input and does not often root ethical frameworks within government system design. This work takes a novel approach in framing ASB intervention as a human-computer interaction problem by embedding an ethical framework into two digital designs, aiming to increase public responsibility and prevent ASB. The first design is extracted from UK public opinion research, the ethical themes include punitive proportionality, personalisation, and responsibility. The second are digital interventions that present a set of QR-based public reporting interfaces and a web-based ASB awareness course that precedes punitive escalation. Our methodology involves structured interviews and online surveys. Results positively evaluated the framework and QR interfaces. Such outcomes could inform the expansion of technological intervention utilisation that does not replace existing punitive approaches, but balances them.

\end{abstract}
\section{Introduction }
According to the Anti-social Behaviour, Crime and Policing Act 2014, Anti-Social Behaviour (ASB) encompasses non-criminal behaviour that causes alarm, harassment, or distress to other individuals outside of the domestic home \cite{asb_act2014}. Examples include hate crime, vandalism, disruptive illicit substance use, and dangerous parking. One million ASB cases were reported in the UK, a result that reoccurred in 2023, 2024, and 2025 \cite{ons2024,ONS2023Crime,ONS2025Crime}. Literature considers it a sociopolitical issue with destructive tendencies, impacting community cohesion and trust in public institutions \cite{millie2007}. Government statutory guidance emphasises ASB enforcement powers through Community Protection Warnings/Notices (CPWs/CPNs), behavioural contracts, spatial orders, fines, and civil injunctions, with the recent 2025 Respect Order expanding police punitive powers \cite{homeoffice_moj_2025_factsheet}.

Existing critique of UK ASB approaches highlights how punitive reliance can deplete  intervention engagement, questioning efficacy\cite{crawford2017}. Additionally, in 2024 a Local Government Association survey highlighted inconsistencies in local powers utilisation, decision-making, and multi-agency information sharing \cite{lga2024}. The limited digitalisation of management tools contributes to inconsistencies, with web-based information, reporting, and police ASB referral forms exhibiting inaccuracies in referral completion, data storage, and information across districts \cite{homeoffice2011_interventions}. This risks resource inefficiency, loss of data, disproportionate decision-making, and public distrust \cite{heap2022}. UK ASB management lacks consistent utilisation of digital systems specifically designed to mediate ethical standards, guide discretionary judgement, encourage public engagement, and offer structured non-punitive interventions. Because of such critique, this research takes a non-punitive approach that focuses on prevention and support for all individuals involved in ASB cases.

Therefore, we developed the following research question: \textit{how can ethical principles be translated into interaction structures that meaningfully influence ASB reporting and prevention?} To explore this, we operationalise ethical mediation through the design of two digital artefacts. Our approach builds on existing HCI work involving ethical mediation, positioning digital systems not merely as a passive tool, but as active structures that shape user behaviour and decision-making. Specifically, we extend PECBR, an ethical guiding framework developed in our previous work, which provides intervention prompts to guide practitioner subjectivity during ASB management \cite{hill2023msc}. The framework is further detailed in sections 1.1 and 3. The first artefact translates PECBR's principles into a QR-based reporting interface designed to guide responsible ASB reporting. The second artefact builds on a collaboration with Swansea City Council, focusing on providing non-punitive intervention for ASB engagers prior to punitive escalation through ASB impact awareness and knowledge training. This paper contributes: 1. an ASB intervention reframing as a HCI problem of ethical mediation; 2. an ethical framework translated into QR reporting interfaces and a digital awareness course; 3. contributions toward an ethically embedded digital interaction approach in public-sector systems. The paper will discuss how the research contributes toward translating ethics into digital intervention design, with contextual considerations outlined.

\subsection{Ethical mediation background}
ASB intervention has rarely been conceptualised as an interaction design challenge. This section discusses how ethical mediation can be applied to digital interventions within ASB management. ASB ethical decision-making is subjectively reliant upon personal discernment of practitioners/police, particularly due to broadly defined ASB parameters\cite{Halford2022ASB}. Notably, there is no standardised ethical framework to guide personal discernment. While subjectivity allows for case personalisation, it also means opportunities for comparative evaluation of intervention strategy, proportionality, and efficiency can be lost.

Our research has been founded on initial informal engagements with local practitioners that helped to establish a research area and scope. This led to inferences that training relies on verbal presentation and 'inherited knowledge' from predecessor professionals rather than standardised technological training. Despite literature calling for better practitioner training \cite{lga2024}, training alone will not guarantee the mediation of ethical subjectivity in ASB decision-making. Yet, due to the government's increasing interest in HCI and Computer Science tools, there is opportunity for this to develop \cite{ukparliament2025_ai}. Specifically, exploring how ethical guidance can be embedded within interface structures and interaction flows.

The PECBR framework (see Appendix), curated from the author's public opinion research, identifies 9 principles to consider in ASB management \cite{hill2023msc}. In practice, the principles function as a framework of evaluative prompts for practitioners and intervention designers. Rather than prescribing fixed outcomes, PECBR supports reflective decision-making by encouraging structured consideration of extraneous variables. To guide ethical practitioner decision-making, the framework identifies the values: Prevention; Personalisation; Empathy; Education; Collaboration; Consideration; Community; Balance; Responsibility. 'Prevention' encourages early intervention and access to support services before escalation. 'Personalisation' and 'Empathy' emphasise case-sensitive evaluation. 'Education' highlights increasing understanding and awareness. 'Collaboration' and 'Community' regards governmental shared responsibility with the public. 'Consideration' stresses accessibility and inclusion. 'Balance' addresses punitive proportionality. 'Responsibility' emphasises engager accountability. While ethical frameworks are valuable, their influence is limited if they remain theoretical or detached from practitioner systems. This paper explores how PECBR's ethical dimensions can influence interface structure, information hierarchy, and guided interaction in digital ASB intervention. Following this, Section 2 situates the work within existing HCI literature; Section 3 details the methodology and evaluation approach; Section 4 presents the results; and Section 5 discusses the research question, related work, findings, limitations, and future directions.

\section{Related work}
This section presents a deep dive of the presented artefacts within three areas of HCI literature: Value Sensitive Design (VSD), civic reporting systems and public-sector interaction design, and behaviour change awareness systems.

\subsection{VSD and ethical frameworks.}
Existing work explores how ethical principles can be embedded within system design, notably through approaches such as VSD and responsible interaction design \cite{friedman1996}. VSD aims to address the ethical impact of technology through integrating human values throughout design processes. It emphasises the consideration of stakeholder values, potential biases, and the broader social implications of technological systems. By embedding human values during design, VSD aims to reduce unintended social harms and improve long-term system effectiveness. Within public-sector systems, such approaches highlight how interface structures and interaction flows may shape how users interpret responsibilities, options, and outcomes. VSD utilisation was successfully applied to service design for prisoner rehabilitation services in Finland and digital governmental collaborative platforms for environmental issues \cite{balcomraleigh2020,sapraz2021}. However applications of VSD often do not examine discretionary intervention systems where ethical judgement and behavioural regulation intersect, including ASB management. This gap is particularly visible in local government enforcement systems, where practitioner discretion strongly influences intervention outcomes. By translating PECBR values generated from public opinion data into interaction structures that mediate practitioner guidance, reporting behaviour, and behavioural awareness, VSD is explored.

\subsection{Civic reporting systems and public-sector interaction design}

QR interfaces have been utilised within research to test public engagement and improve communication, with results inferring technology can be integrated within ethical governmental campaigns \cite{goggin2024,Lorenzi2014QR}. Within HCI and civic technology research, QR-mediated interfaces are explored as low-friction entry points between public information systems and citizen interaction \cite{Lorenzi2014QR}. During the COVID-19 pandemic, governments internationally deployed cost-effective QR codes for various public communication systems \cite{goggin2024}. These systems effectively connected physical locations, individual devices, and digital databases through simple scanning interactions. Deployed in Australia, a mobile application for reporting city maintenance issues through geo-tagged photo reports to the user's local government was successfully engaged with, inferring public familiarity with scanning interfaces may determine efficacy  \cite{foth2011}. The interface reduced friction in public reporting and bettered relationships between the public and local governments, demonstrating how interface structure can shape participation in civic reporting systems. Additionally, research suggests that the success of local government QR systems depends on QR literacy, public receptiveness, consistency, and economic resources \cite{foth2011,fauzia2025,fogg2009}. If informed by ethical frameworks such as PECBR, QR interfaces have the potential to improve ASB reporting, community responsibility, and awareness. The QR interfaces presented in this paper demonstrate how embedding ethical prompts within reporting interface structure influences how responsibility and reporting is interpreted by users, with survey feedback informing design.

\subsection{Behaviour change and awareness interventions in HCI}

Digital systems are frequently designed to influence user behaviour through structured interaction design. Foggs' FBM model discusses how behaviour is a product of motivation, ability, and triggers, which may inform the design of HCI technologies to achieve engagement \cite{fogg2009}. Behaviour-change technologies in HCI commonly employ prompts, feedback loops, graphical user interfaces (GUIs), and interactive learning tools to support behavioural reflection. Such systems aim to create frictionless interaction flows that guide users toward specific learning outcomes or behavioural changes \cite{Akmal2021Spoonful}. Research outlines how web-based platforms can address concerns of inclusivity, accessibility, and transparency when involving the public in environmental intervention and communication with practitioners \cite{khan2025}. It demonstrates how structured online platforms can support communication, reflection, and collaborative decision-making through guided interaction design.

To elaborate, persuasive technologies (city-sponsored websites and mobile applications) have been utilised to elicit societal behavioural change regarding global carbon emissions and environmentally responsible behaviours in Northern America \cite{corbett2022}. These technologies utilised the persuasive systems design (PSD) framework design principles concerning task, social, and integration support \cite{OinasKukkonen2009PSD}. However, while behaviour-change technologies are widely applied in domains including environmental sustainability, public awareness, and healthcare, they are rarely explored within practitioner discretionary decision-making \cite{Adaji2022PersuasiveTech}. The following artefacts explore how behaviour-change and persuasive interaction design can be applied to ASB digital intervention. For instance, the digital awareness course is designed to be a behavioural change technology, influenced by persuasive systems design to provide engagers with opportunities to reflect and change disruptive behaviour.  
\section{Methodology}

This research adopts a design-oriented HCI methodology combining ethical framework development with iterative artefact design and exploratory evaluation. In this section, we outline the procedures and planned procedures for PECBR, the QR reporting interfaces, and the web-based awareness course. Then, existing empirical analysis and future empirical data collection is outlined. All studies were approved by Swansea University's Science and Engineering ethics committee.

\subsection{Development of the artefacts} 

\textbf{The PECBR framework.} An online anonymous survey study collated quantitive and qualitative data on ASB knowledge, understanding, and opinion through the online survey platform 'Jotform'\cite{hill2023msc}.  58 non-experts were recruited in the UK through social media over one week and tasked with suggesting interventions to three hypothetical case studies. Results were analysed thematically using Braun and Clarke's reflexive thematic analysis (RTA) \cite{Braun2023ReflexiveTA}. PECBR (see Appendix) was visually curated in the design platform Canva, with primary colours blue and black chosen in consideration of accessibility needs of colour visual impairments\cite{jenny2007}. Theme descriptions were based on familiarisation with the data, codes generated during RTA, and local council approaches.

\textbf{Evaluation of PECBR.} PECBR was first evaluated during an interview study with one local ASB practitioner. The sample number was due to the exploratory nature of this evaluation and limited sampling pool. The study presented the participant with quantitive questions surrounding their procedures, for instance, 'Do you feel ASB practitioners would benefit from a national ethical framework for ASB decision-making?'. The participant was then presented with three hypothetical ASB case studies and asked to provide interventions. The PECBR model was presented alongside a fourth hypothetical case study, where they were asked to present intervention using the framework as reference and asked questions surrounding PECBR's use in practice. Qualitative results were analysed using Braun and Clarke's RTA and comparative analysis. 

\textbf{QR interface design and evaluation.} The QR-based reporting interfaces were informed by the PECBR values when considering the guided reporting pathway features. Specifically, PECBR values align with the aims of: improving ASB knowledge engagement (education value), easing accessibility to ASB information reporting through QR interfaces (consideration value), and community member accountability through increasing reporting (community and responsibility values). Posters were created through the design platform Canva. 

\begin{figure}[htbp]
\centering
\includegraphics[width=0.82\textwidth]{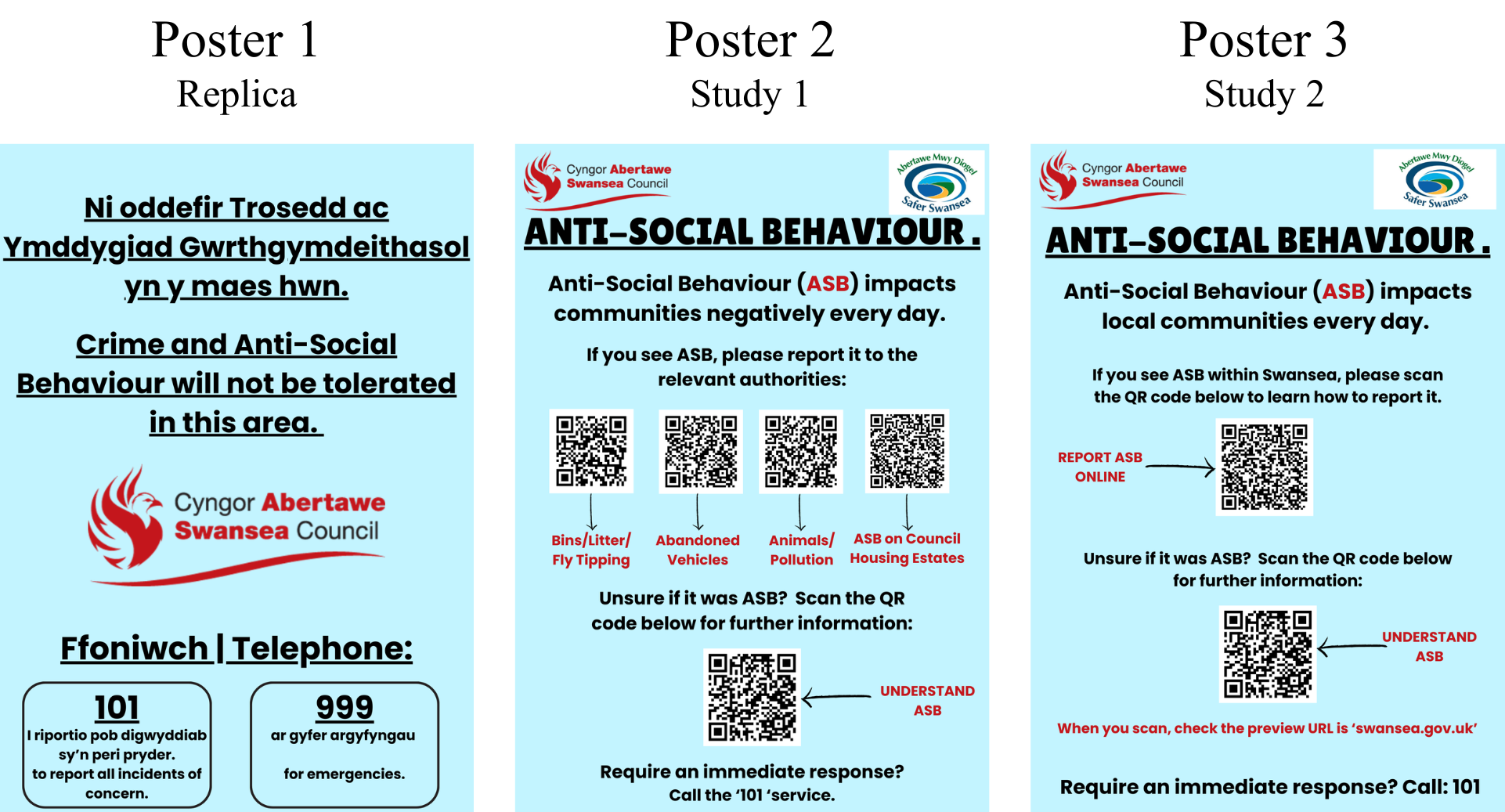}
\caption{QR reporting interfaces 1,2 and 3.} \label{fig:posters}
\end{figure}

Poster 1 is a replica of the original local government poster found in Swansea, with the baseline colour altered from yellow to blue for visual impairment accessibility consideration. Poster 2 was informed by local ASB practitioners regarding helpline details and representative logos. Recognisable council domain indicators were utilised to strengthen user trust and reduce hesitation when interacting with the interface. Static QR codes were generated on Canva using the council URLs, and cognitive load was considered by using succinct text. Posters 1 and 2 were compared in an anonymous online study (Poster study 1) created on Jotform. 57 non-experts were recruited through poster advertisement (created on Canva) distributed at Swansea University and on social media. Participants were asked to assess Poster 1 and provide feedback, then assess and engage with Poster 2's QR codes and provide feedback. Quantitive data was analysed statistically, with percentage averages of numerically scaled answers calculated.

Poster 3 was created after Poster Study 1 identified pain points in the interaction design. The local council webpages were assessed by utilising these user interaction pain points and interaction flow tests. The statistical insights from study 1 provided the evidence required to authorise changes to the local council website, including reducing the number of reporting webpages to one. Further alterations (information clarity) were assigned to the local council website development team to create a streamlined QR reporting interaction flow to prevent friction and depleted motivation to engage. Two temporary static QR codes were generated on Canva. A prompt to check preview URLs of scanned QR codes was introduced. Blue was the primary design colour for consideration of colour visual impairments as blue does not often shift when contrasted with black text \cite{jenny2007}. Additionally, Welsh translated versions were created for inclusivity, not pictured.

\textbf{Awareness course design.} The web-based awareness course was designed on Microsoft PowerPoint as a low fidelity prototype (see figure 2). It aims to become a digitally coordinated behavioural change technology designed through Laravel, a web development framework, that decreases user motivations to engage with ASB prior to punitive escalation. The artefact implicates PECBR's themes of: 'Education', the modules teach ASB impacts and definitions; 'Responsibility', engagers show accountability through course engagement; 'Balance', the course is a less punitive approach delivered with punitive CPNs; 'Prevention', the course aims to reduce engager motivations to re-engage with ASB. 

\begin{figure}[htbp]
\centering
\includegraphics[width=0.7\linewidth]{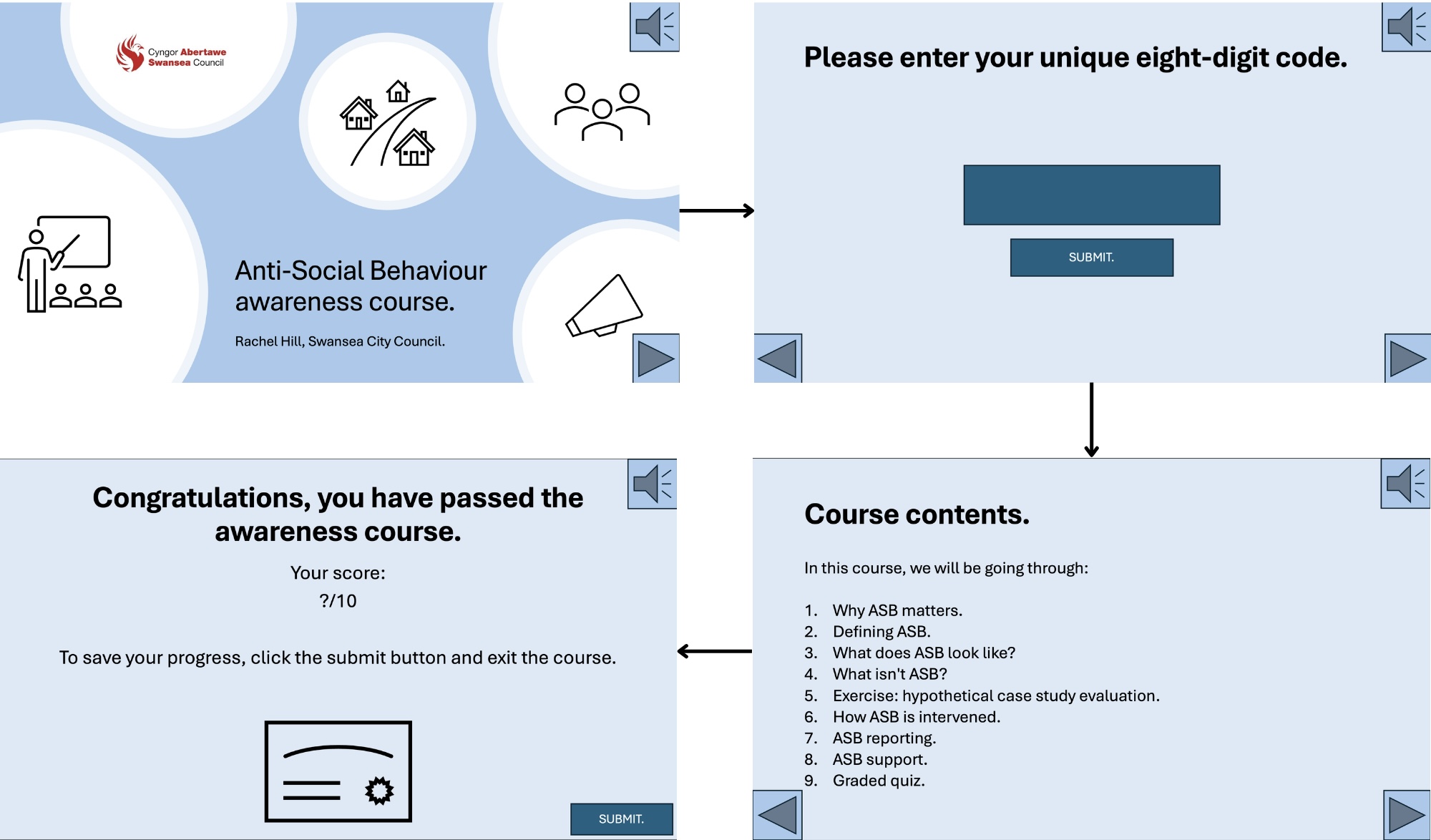}
\caption{ASB awareness course low-fidelity prototype slides 1,2,3 \& 35.} \label{fig:awarenesscourse}
\end{figure}

The prototype includes graphical user interfaces (GUIs) such as navigation buttons, 'read aloud' functions, and adjustable subtitle functions to increase accessibility to the course (see figure 2). The 'read aloud' functions were tested within the draft design through artificially generated MP3 audio files of the module text. MP3 files were converted into WAV. files and uploaded to Microsoft Powerpoint. A demo of the 'read aloud' functions were informally tested with Swansea practitioners. The primary design colour was blue contrasting with black, as research recommends this palette for colour visual impairments. \cite{jenny2007}

Local government/Home Office statutory ASB guidance and local ASB practitioners informed module content, with information divided to prioritise efficient cognitive load and information hierarchy \cite{homeoffice2025_powers,Hollender2010CLTHCI}. The module sequencing is as follows: Landing page; Unique identification code login; Introduction; Module 1: Why ASB matters; Module 2: Defining ASB; Module 3: What ASB looks like; Module 4: What is not ASB; Module 5: Hypothetical case study evaluation; Module 6: How ASB is managed; Module 7: Reporting ASB; Module 8: ASB support; Module 9: Graded quiz (10 questions); Notification option one: Pass (70\% minimum grade)- course completion. Notification option two: Fail- course restarts. Personalised features include practitioner control of modules and support service recommendation per case, accessed through an administrator login. For engagers, unique access codes (8 digits randomly generated) are provided by letter alongside CPNs. Courses will be delivered through computers operating in public libraries at designated times with practitioners present to supervise. 

\section{Results}
In this section empirical evaluation for PECBR, Poster 1, and Poster 2, and evaluation strategies for future work is outlined. 

\textbf{PECBR framework evaluation.} The evaluation of the PECBR framework, based on RTA, generated 10 themes including: multi-agency approach, engager history, subjectivity; flexibility; intervention receptiveness; punitiveness; equitable support; community; shared responsibility; ethical intervention. Comparative analysis indicated that 9 out of 10 identified themes aligned with 8 of the 9 PECBR values, suggesting a strong conceptual correspondence between practitioner reasoning and the proposed ethical framework. The participating practitioner also reported that PECBR was useful in guiding intervention decisions, indicating its potential applicability to real-world ASB contexts. This evaluation approach has been extended through an ongoing study design involving practitioners across Wales to further assess the framework's generalisability.

\textbf{QR interface evaluation (Poster Study 1).} The statistical analysis of the 57 participant responses revealed the following results (see table 1). With 82\% of participants having witnessed ASB, but 72\% never reporting it, a significant gap is highlighted between awareness and reporting behaviour. Engagement with Poster 1 (baseline design) resulted in limited knowledge improvement, with 49\% of participants indicating no increase in understanding of ASB reporting processes. In contrast, Poster 2  (PECBR-informed design) demonstrated a substantial improvement, with 96\% of participants reporting increased knowledge after exposure. Additionally, 70\% of participants reported that the QR-based interaction was easy to use, suggesting QR integration can effectively reduce barriers to reporting. However, 16\% of participants recommended reducing the number of QR codes, indicating potentially high cognitive or visual load in the interface design. These findings informed the iterative development of Poster 3. 

\begin{table}[htp!]
\centering
\caption{Poster study 1 responses for ASB poster/QR engagement.}
\label{Poster 1 statistical analysis}
\begin{tabular}{l c}
\hline
\textbf{Survey item} & \textbf{Participant percentage (\%)} \\
\hline
Have witnessed ASB & 82 \\
Never reported ASB & 72 \\
ASB knowledge did not increase from Poster 1 & 49 \\
ASB knowledge increased from Poster 2 & 96 \\
Found Poster 2 QR codes easy to use & 70 \\
Recommended decreasing QR code frequency on Poster 2 & 16 \\
\hline
\end{tabular}
\end{table}

\section{Discussion}
Our findings demonstrate that ethical frameworks such as PECBR can be operationalised within HCI design to influence user behaviour and decision-making in ASB contexts. Through embedding PECBR within QR reporting interfaces and digital awareness intervention, this work moves ethics from abstract policy guidance into tangible digital tools. 

The QR reporting interfaces present how language framing and information entry points shape public reporting, understanding, and accountability. The Poster Study 1 early empirical results showed significant increase in user knowledge after engaging with poster 2's QR interface. This infers credibility in QR reporting interfaces for ASB informed by ethical frameworks. Moreover, 70\% indicated the QR codes were easy to use. Accompanied by discussed research on public familiarity with QR interfaces influencing efficiency, the results suggest they are an effective intervention to introduce within ASB management \cite{goggin2024}.

However, the deployment of such systems is subject to practical and regulatory constraints. Detailed planning and communication between governing bodies must take place prior to QR reporting interface deployment. In Wales, existing restraints within highway laws, local government guidance and powers, public lawsuits, and property ownership can slow, or cease, deployment \cite{govwalesTraffic2026,govwalesWelshLanguage2026}. Utilising QR interfaces as digital public engagement tools is a financially effective method in ASB management, but deployment takes time. Additionally, safeguarding in design must be considered, as 'quishing' (QR code scamming) is a risk \cite{musa2025,govwalesQR2026}. False QR codes can be placed over existing codes to deface and lead users to malicious websites \cite{rys2025}. In Poster 3, users were warned to check URL previews upon scanning to avoid 'quishing'. Utilising mounted frames and controlling poster height are also recommended to avoid malicious defacement. Monitoring of interfaces is possible yet reliant on local government resources. 

The PSD framework outlined guiding design principles for interaction design, in particular highlighting the importance of information integration (sign posting) within design \cite{corbett2022}. This has influenced the feature of tailored support services provided in the awareness course depending on administrator discernment. Moreover, the FBM model is utilised in the course design, as functions aim to ensure motivation (users engage to avoid punitive escalation), ability (engagement is ensured through accessible public library deliverance), and triggers (users are notified when CPNs are delivered to their home address) \cite{fogg2009}. Additionally, the application of behavioural change technologies designed to intervene with social behaviour of governmental concern will always be restricted by resources, which depends on various factors. Moreover, to create technologies that are accessible to all individuals of varying economic stability requires publicly accessible resources, such as public libraries, which is what the awareness course intends to utilise. Whilst this would work on a locally known basis, this cannot be guaranteed nationally. Therefore, financial and economic factors must be considered during deployment design of digital ASB interventions to ensure accessibility.

\subsection{Limitations and future work}
\textbf{Limitations.} The sample size of PECBR's evaluation was limited due to the local ASB team consisting of one professional. This is a common case across the UK, as financial restraints in local government often means ASB teams are minimised. To address this, wider practitioner evaluation in Wales is underway. It is recommended that future work in this field considers the constraints upon Local Government that cause limited practitioner sampling pools. Moreover, all samples were collected in the UK only. Though ASB is a worldwide issue, legislation differs cross-culturally which impacts intervention deployment possibilities. ASB itself is vaguely defined and takes differing forms depending on culture and context \cite{Halford2022ASB}. Therefore, this research could be generalisable cross-culturally once adjusted to and tested within these variables. Regarding design, colour visual impairments were considered, however the red details on Posters 2 and 3 may shift to a dark brown with certain colour visual impairments \cite{jenny2007}. Therefore, future variant designs should consider colour vision deficiency accessibility.

\textbf{Future work.} Pending updates to a local council webpage, poster 3 is awaiting deployment. Once finalised, dynamic QR codes will be generated for the new URLs to allow for empirical data collection of engagement frequencies. Webpage engagement will be monitored by the council through Google analytics. This enables multilayered empirical observation while maintaining low-friction public access to ASB reporting systems. The ASB awareness course low fidelity prototype will be made into a working prototype on Laravel. It will evaluated by ASB engagers in a structured workshop usability study with an experimenter present. In this formative mixed-methods user study, quantitive data on the reported ASB will be collected through a preliminary survey, then the web-based awareness course will be tested, followed by quantitative/qualitative evaluation.

\section{Conclusion}
This paper demonstrates how ethical principles can be operationalised within HCI design for public-sector systems. Through the PECBR framework and its translation into QR-based reporting interfaces and a web-based awareness course, ethics are embedded within interaction structures, moving away from abstract guidance. Our findings suggest that ethically informed design can improve user understanding and engagement in ASB reporting. While exploratory, these results indicate the potential for ethical mediation being a practical design approach. It is suggested that future work considers inputting ethical frameworks in HCI design for governmental deployment, to ensure that generated outputs reflect the human values and experiences of users, increasing engagement and long-term efficiency.

\section*{Acknowledgments}

This work was funded by EPSRC grant number EP/S021892/1 and Swansea City Council.

\newpage
\section{Appendix}
\begin{figure}[htbp]
\centering
\includegraphics[width=0.8\linewidth]{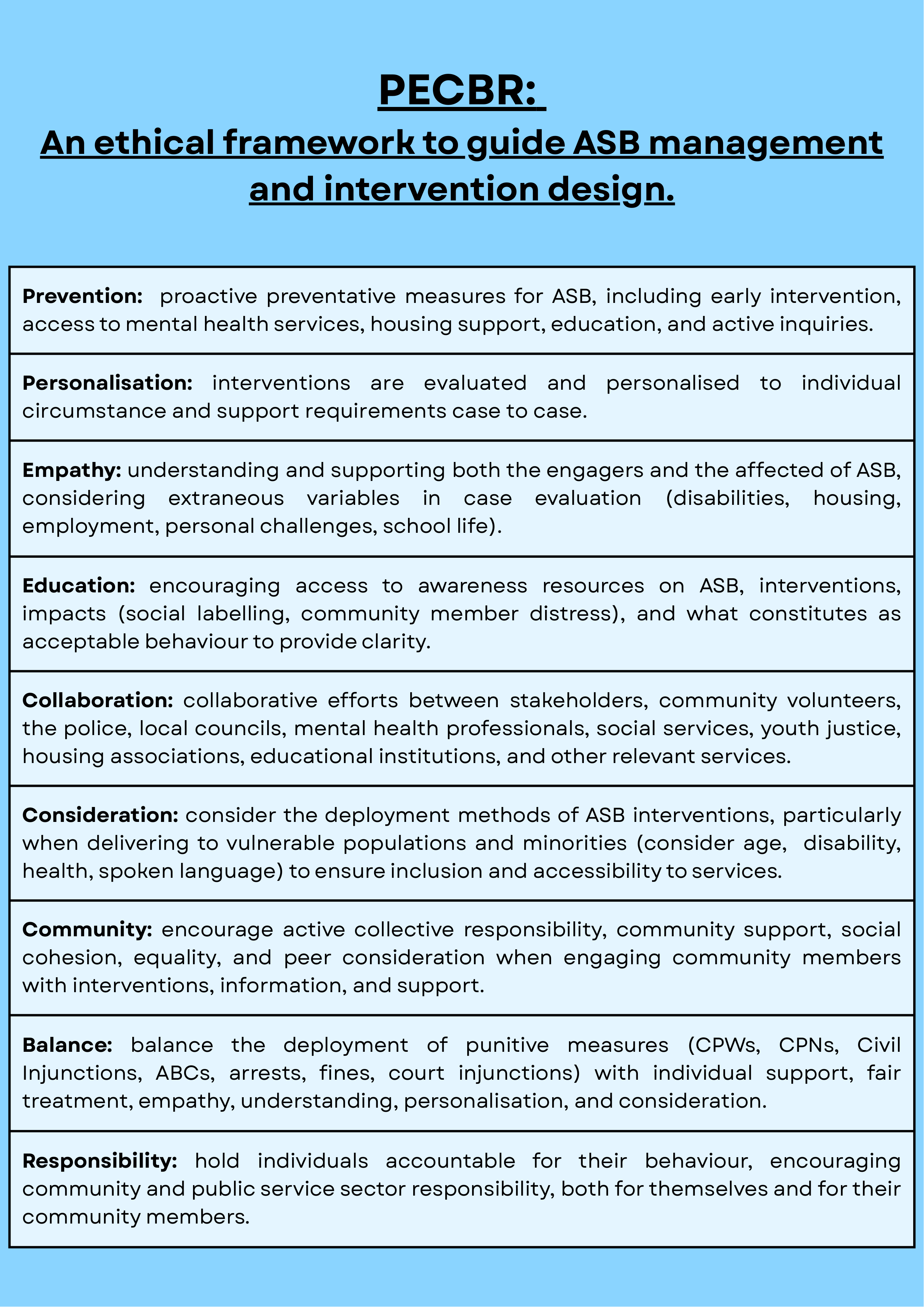}
\caption{The PECBR framework, Hill, R (2023).} \label{fig:pebcr_framework}
\end{figure}

\newpage
\bibliographystyle{splncs04}
\bibliography{refs}
\end{document}